\documentstyle[11pt,epsfig]{article}
\newcommand \be{\begin{equation}}
\newcommand \ba{\begin{eqnarray}}
\newcommand \ee{\end{equation}}
\newcommand \ea{\end{eqnarray}}

\begin{document}

\begin{center}
{\LARGE The US 2000-2003 Market Descent: Clarifications}
\end{center}
\bigskip
\begin{center}
{\large Didier
Sornette {\small$^{\mbox{\ref{igpp},\ref{ess},\ref{lpec}}}$} and Wei-Xing Zhou
{\small$^{\mbox{\ref{igpp}}}$}}
\end{center}
\bigskip
\begin{enumerate}
\item Institute of Geophysics and Planetary Physics, University of California,
Los Angeles, CA 90095\label{igpp}
\item Department of Earth and Space Sciences, University of
California, Los Angeles, CA 90095\label{ess}
\item Laboratoire de Physique de la Mati\`ere Condens\'ee, CNRS UMR 6622 and
Universit\'e de Nice-Sophia Antipolis, 06108 Nice Cedex 2, France\label{lpec}
\end{enumerate}

\begin{center}{\today}\end{center}

\begin{abstract}
In a recent comment [Johansen A 2003 An alternative view, Quant.
Finance 3: C6-C7, cond-mat/0302141], Anders Johansen has
criticized our methodology and has questioned several of our
results published in [Sornette D and Zhou W-X 2002 The US
2000-2002 market descent: how much longer and deeper? Quant.
Finance 2: 468-81, cond-mat/0209065] and in our two consequent
preprints [cond-mat/0212010, physics/0301023]. In the present
reply, we clarify the issues on (i) the analogy between rupture
and crash, (ii) the Landau expansion, ``double cosine'' and
Weierstrass-type solutions, (iii) the symmetry between bubbles and
anti-bubbles and universality, (iv) the condition of criticality,
(v) the meaning of ``bullish anti-bubbles'', (vi) the absolute
value of $t_c-t$, (vii) the fractal log-periodic power law
patterns, (viii) the similarity between the Nikkei index in
1990-2000 and the S\&P500 in 2000-2002 and (ix) the present status
of our prediction.
\end{abstract}

In a recent comment \cite{J03QF}, Anders Johansen has criticized
our methodology and has questioned several of our results
published in this journal \cite{SZ02QF} and in our two consequent
preprints \cite{ZS02,ZS03}. We regret the controversial tone
adopted in \cite{J03QF} but welcome this opportunity to clarify
our work further.

In a series of works starting with \cite{1996crash} (see
\cite{Sor03Book} and references therein), financial bubbles have
been defined as regimes in which the stock market exhibits an
unsustainable super-exponential growth, that can be characterized
quantitatively as a genuine critical phenomenon with specific
log-periodic power-law (LPPL) signatures. The underlying mechanism
is proposed to be found in imitation between investors and their
herding behavior, which lead to self-reinforcement positive
feedbacks.

In \cite{NikkeiPred}, Johansen and Sornette introduced the concept
of ``anti-bubbles'' to describe decaying LPPL price trajectories
that are sometimes found to follow very large market highs. Based
on models of imitation between investors and their cooperative
herding behavior \cite{NikkeiPred,Idesor1,Idesor2}, it was
realized that speculation and imitation also occur during bearish
markets, leading to price trajectories that seem approximately
symmetric to the accelerating speculative bubbles ending in
crashes, under a time reversal transformation ($t_c-t \to t-t_c$
where $t_c$ is a critical time corresponding to the end of the
bubble or the start of the anti-bubble). The two first examples of
anti-bubbles were found \cite{NikkeiPred} in the Japanese Nikkei
stock index from 1990 to 1998, whose analysis led to the
successful prediction of two trend reversals
\cite{evaljapan,Sor03Book}, and in the Gold future prices after
1980, both after their all-time highs. Several other examples have
been described in the Russian stock market \cite{JSL99} and in
emergent and western markets \cite{emergent}. Our recent work
\cite{SZ02QF,ZS02,ZS03} adds many other cases that all started in
the summer of 2000.

\vskip 0.3cm {\bf Status of the rupture analogy}. More precise and
probably more relevant than the analogy with material rupture is
the concept of a finite-time singularity as developed in
\cite{Idesor1,Idesor2,fts,gs}, which emerges from positive
feedbacks. The concept of a finite-time singularity is the
counterpart in the time-domain of the concept of criticality. The
fight between positive and negative feedbacks is the key concept
underlying the proposal of LPPL signatures of speculative bubbles
and anti-bubbles in stock markets \cite{Sor03Book}.

\vskip 0.3cm {\bf Landau expansion, ``double cosine'' and
Weierstrass-type solutions}. A. Johansen criticizes our use of the
``double cosine'' function on the basis that a sound theoretical
justification is lacking, while he puts his faith in the Landau
expansion introduced in \cite{SJ97} and extended up to third order
in \cite{NikkeiPred}. Actually, the full solution of the simplest
renormalization group equation for a critical point has been
analyzed in depth in \cite{GS02} and provides an improvement of
these approaches in the form of Weierstrass-type functions of the
form
\begin{equation}
\ln [p(t)] = A + B \tau^{m} + {{\Re}}\left(\sum_{n=1}^N C_n
{\rm{e}}^{i\psi_n}\tau^{-s_n}\right)~, \label{Eq:lnpt}
\end{equation}
where $\tau=|t-t_c|$, $s_n = -m+in\omega$ and $\Re()$ is the real
part operator. The existence of different phases $\psi_n$ incriminated
by A. Johansen can be seen to derive naturally from the Mellin transform
of the regular part of the renormalization group equation. In simple
words, the different phases $\psi_n$ embody an information on the
mechanisms of interactions between investors. There
is thus a sound theoretical justification for such a phase shift
(understand $\psi_2-\psi_1$) between the first and the second harmonics
(understand the first two terms $n=1$ and $2$ of the expansion
(\ref{Eq:lnpt})). When the phases have certain relationships (phase
locking), a discrete hierarchy of critical times emerge, which has been found
to describe very well the US stock market since the summer of 2000
\cite{ZS03}. A. Johansen's misconception can probably be traced to the
incorrect idea that the phase of the simple cosine
formula (case $N=1$) has no financial meaning because it can be
gauged away in a redefinition of the time scale.

\vskip 0.3cm {\bf Symmetry between bubbles and anti-bubbles and
universality}. From a mechanistic view point, we advocate the
existence of anti-bubbles from the idea that the fight between
positive and negative feedbacks is operative both in bullish as
well as in bearish markets \cite{Sor03Book}. From a descriptive
view point, our recent works  \cite{SZ02QF,ZS02,ZS03} just follow
Johansen and Sornette's previous works
\cite{NikkeiPred,evaljapan,JSL99,emergent}, which introduced the
concept of an ``anti-bubble'' from a symmetry perspective. A
symmetry may have distinct consequences. It can be used to justify
the same functional expressions both for bubbles and anti-bubbles.
Thus, in the mathematical expressions, the symmetry between
bubbles and anti-bubbles amounts to changing $t_c-t$ for bubbles
to $t-t_c$ for anti-bubbles. Here, we should stress that, if a
LPPL anti-bubble follows a LPPL bubble (which is not the general
case), the critical time $t_c$ is not generally the same. A
noteworthy exception is the Russian stock market around 1997
\cite{JSL99}. We report in \cite{SZ02QF,ZS02} dozens of
anti-bubbles in many different stock markets worldwide which
started almost all in August 2000, that is, 4 months later than
the end of the ``new economy'' bubble and its crash in April 2000.
Another case is Chile around 1994-1995, where the bubble ended in
February 1994 while the anti-bubble started in July 1995
\cite{emergent}.

A. Johansen would like that the symmetry between bubbles and
anti-bubbles should be extended so that the same log-periodic
angular frequency $\omega$ describes both cases. He thus invokes
more than just a functional but a numerical symmetry. We think
that this belief may be too rigid at the present time when we
still have a rather limited understanding of this complex problem.
We propose an open-minded approach more adapted to a learning
phase. It is correct that, for LPPL bubbles, there is rather
well-defined cluster of values for $\omega \approx 6.36 \pm 1.56$
and for $m \approx 0.33 \pm 0.18$ as reported in \cite{JS03} (see
equation (4) of \cite{Joh03}). For anti-bubbles in the USA S\&P
and in many EU markets, we find almost the same value $\omega
\approx 12$. This value is comparable with those obtained for the
anti-bubbles in the Latin-American markets and Western markets in
the 1990's \cite{emergent}. It is interesting that this value
$\omega \approx 12$ is approximately twice the most probably value
$\omega$ found for LPPL bubbles. Does it correspond to a
log-periodicity different from that of bubbles? Probably not for
the following reason: we have found in \cite{SZ02QF,ZS02} that
both $\omega$ and $2 \omega$ were quite significant in the
anti-bubbles, including the Nikkei case that started in 1990. Due
to the probable variation of the strength of nonlinear processes
in the stock markets, it can be expected that the amplitudes of
the first and second harmonics can be different from one
realization to the next. Within the renormalization group
framework, the relative strength of the first and second harmonics
is controlled by the regular part \cite{GS02} which describes the
specific interactions of the investors that led to a given
realization of the market. Let us add that the importance of the
role of log-periodic harmonics has been demonstrated for
turbulence \cite{turb1,turb2}, where the evidence is much
stronger. For the emergent markets, the LPPL signatures are not as
significant as for the major western markets, as noted already in
\cite{emergent}. A. Johansen also notes the $\omega$'s found for
different worldwide markets are not peaked and may be due to
noise. Instead, we think that this is due to a possible lack of
sufficient robustness of the fits, which does not diminish the
evidence for log-periodicity but suggests to interpret with care
the specific reported values. This can be seen from the fact that,
if we impose the additional condition in our fits that the
different worldwide markets exhibit an anti-bubble with the same
critical time $t_c$, we find that their angular log-periodic
frequency $\omega$ are very close to each other. The
quasi-simultaneity of the starting time and the ensuing strong
synchronization of the anti-bubbles exhibited by the major stock
markets in the world, which has been documented in \cite{ZS02},
provides an additional justification for the use of the same
critical time $t_c$.

\vskip 0.3cm {\bf Criticality}. A. Johansen criticizes our
abandoning of the constraint $m<1$ as a necessary condition to
qualify the existence of a bubble or anti-bubble, suggesting that
we have renounced the concept of criticality. There are several
issues here that need to be distinguished. First, our many tests
performed by the present authors and previously by A. Johansen
with D. Sornette (reported as the work that Johansen performed
with Matt Lee in \cite{J03QF}) show that the condition on the
exponent $m$ is much less effective in the detection of bubbles
than a condition on $\omega$ for instance (see also discussion in
Chapter 9 of \cite{Sor03Book}). This is one justification for
abandoning any constraint on this rather sensitive parameter
to ``let the data speak.'' Second, finding values of $m \geq 1$
does not amount to an absence of criticality, because the equation
is still critical (that is, it exhibits
a singularity) due to the presence of the theoretically
infinite hierarchy of log-periodic oscillations. In other words,
criticality remains present due to the imaginary part $\omega$ of
the exponent $s_n = -m+in\omega$ of the LPPL (see equation
(\ref{Eq:lnpt})) as long as it is non-zero, whatever the value $m$
of its real part. Third, we can relax the condition $0 < m < 1$
for the present purpose because our LPPL formulas describe only a
finite range of the time interval: it is well-known that true
singularities do not exist in nature as friction, finite-size
effects and other regularization mechanisms come into play close
enough to the theoretical mathematical singularity. What is
important is the ability of the LPPL formula to describe with good
accuracy a large range of the data, not necessarily the very close
proximity to the phantom singularity. In this respect,
we refer to the rather detailed discussion of the effect of finite-size
effects on singularities presented in \cite{fts}.

\vskip 0.3cm {\bf ``Bullish anti-bubbles''}. In our analysis of
the largest stock markets in the world, we have identified six
examples which give a positive coefficient $B$ in (\ref{Eq:lnpt}).
In particular, the statistical significance of this result is very
high for Australia, Mexico and Indonesia. This regime $B>0$ is
different from the normal bubble and anti-bubble cases previously
reported for which $B<0$. This regime $B>0$ has been coined
``bullish anti-bubbles'' \cite{ZS02} to describe the joint
features of decelerating log-periodic oscillations and of an
overall increasing price. In contradiction with Johansen's remark,
this regime $B>0$ does not lead to infinite prices in a finite
time but describes a long-term growth which turns out to be slower
than standard exponential growth. The same remark applies for
$m>1$.

\vskip 0.3cm {\bf Absolute value of $t_c-t$}. In complete
disagreement with A. Johansen's remark, our use of $|t_c-t|$ in
our fits to locate the critical time $t_c$ does not abandon
``another restriction coming from the data.'' Rather than adding a
degree of freedom, this approach instead removes an arbitrariness
previously present in the fitting procedure in choosing the time
interval over which the fit is performed. Rather than determining
an approximate starting time and/or estimating the critical time
$t_c$ by the location of the largest market peak, using  $|t_c-t|$
makes the fits almost independent of the chosen starting time.
This improved robustness has been documented in details by our
many numerical tests presented in \cite{SZ02QF,ZS02}.

\vskip 0.3cm {\bf Fractal LPPL patterns}. As we quoted in
\cite{SZ02QF}, Drozdz et al. \cite{Drozdz} have reported the
existence of LPPL within LPPL within LPPL, using eye-balling in a
single case. As mentioned by A. Johansen \cite{J03QF}, he with D.
Sornette studied this phenomenon rather systematically about a
year earlier but never published due to the rather marginal
quality level of the results. In \cite{SZ02QF}, we mentioned that
the worldwide anti-bubble started in the summer of 2000 has also
left its imprint on the Japanese market, leading to an anti-bubble
within the large scale anti-bubble that started in January 1990.
This possibility of structures within structures is expected on
general grounds from the renormalization group model of LPPL
singularities leading to Weierstrass-like solutions (see
\cite{ZS03,Sor03Book}). The problem is that such observation is
not very robust when one
goes to small time scales, probably due to the fact that ``noise''
and idiosyncratic news affect more and more strongly the price
time series, the smaller is the time scale of observation.
However, we note that our report \cite{SZ02QF} of a $2.5$ year
long anti-bubble decorating a $13$ year long anti-bubble of the
Nikkei index should have a special status because both time scales are
sufficiently long to compare with the time span over which
previous LPPL have been qualified. A. Johansen himself
acknowledges that ``the real success was with a LPPL analysis on
time scales of one to two years of data.'' Our report in
\cite{SZ02QF} passes this criterion and should thus be considered
at a level different from the published \cite{Drozdz} and
unpublished analyses at smaller time scales.

\vskip 0.3cm {\bf Similarity between the Nikkei index in 1990-2000
and the S\&P500 in 2000-2002}. Johansen downplays the ``remarkable
similarity'' we as well as many observers noticed between these
two markets. First, the factor of $2$ in the value of the
log-periodic frequency is explained by the competition between the
two first harmonics $n=1$ and $n=2$ in (\ref{Eq:lnpt}), as we
explained above. In \cite{SZ02QF}, we stress the remarkable similarity
in the two markets with respect to the existence of two harmonics in both cases.
Second, the Nikkei did go through a now
well-recognized speculative bubble culminating at the end of
December 1989, even if its price trajectory does not qualify as a
very good LPPL. We note in this vain that an anti-bubble is
usually the follow-up of very high prices, not necessarily of a
LPPL bubble. Even in the case of the US market, we stressed above
that the critical time of the bubble occurred $4$ months before
the critical time of the following anti-bubble. This
again stresses that one should exercise caution in twinning
rigidly in time the occurrence of bubbles and of anti-bubbles.
Third, Johansen argues that the analysis of the Nikkei was based
on $9$ years of data compared with less than $3$ years for the US market
which, he argues, makes these two cases apart.
Johansen forgets to mention is that the $9$ years of Nikkei data
required the use of a log-periodic formula extended to third-order
in the Landau expansion mentioned above while the analysis in
\cite{SZ02QF} of the US market used only the first-order formula
and its extension with a second harmonics. Johansen and Sornette's
initial analysis of the Nikkei data in \cite{NikkeiPred} showed
that, similarly to the US market, the first three years of the
Nikkei time series could be adequately described by the
first-order formula. It is by extending to large time horizon that
it was necessary to use the higher-order terms in the Landau
expansion. It is also interesting to note that there was a global
anti-bubble starting in January 1994 in the major western stock
markets \cite{emergent}, which also bears similarities to the
present worldwide 2000-2003 anti-bubble case \cite{ZS02}. The
global anti-bubble in the mid-1990's lasted less than one year,
while the 2000-2003 anti-bubble is still alive on many more
markets, resulting in a much higher statistical significance
level. There is thus no qualitative nor quantitative difference
between the Japanese and USA data sets. We would like to add
that the similarity between the Nikkei in 1990-2000 and the
S\&P500 in 2000-2002 can be further strengthened by paralleling
the economic and financial distresses of the two countries, as
explained in \cite{SZ02QF}.

\vskip 0.3cm {\bf Status of the prediction}. Finally, ``The proof
of the pudding is in the eating.'' The ultimate evidence attesting
the true nature of something lies in the verification of ex-ante
predictions by future data. We have offered the predictions for
the future of the US market in \cite{SZ02QF,ZS03} and of many
worldwide markets in \cite{ZS02} as an important additional step
for testing the LPPL hypothesis. These predictions for the S\&P500
US market are compared with the realized values and are also
updated monthly (go to the URL
http://www.ess.ucla.edu/faculty/sornette/ and then click on ``The
future of the USA stock market''). Recall that the prediction
published in \cite{SZ02QF} was made at the end of August 2002. At
the time  of the latest comparison, March 18 2003, one can see
that we correctly predicted the recovery of the market until the
end of 2002 but missed the severe drop that followed, which was
probably due to the uncertainties associated with the coming war
with Iraq. We should also stress that these last months have
exhibited a very large volatility, leading to deviations from our
prediction that are however comparable in magnitude with previous
deviations in the in-sample period. Our predictions are
fundamentally ``low-frequency'' in nature and cannot obviously
capture the detailed idiosyncratic volatility. The comparison
between our predictions and the realized price should thus be made
at the time scale of the prediction horizon, that is, from August
2002 till summer 2004. We stand by our prediction that the market
should appreciate somewhat and then resume its overall bearish
anti-bubble descent.

\bigskip
{\bf Acknowledgments:} This work was partially supported by
the James S. Mc Donnell Foundation 21st
century scientist award/studying complex system.

\end{document}